\documentclass[fleqn,usenatbib,useAMS]{mnras}

\usepackage{graphicx}	
\usepackage{amsmath}	
\usepackage{multicol}        
\usepackage{bm}		
\usepackage{pdflscape}	
\usepackage{latexsym}
\usepackage[authoryear]{natbib}
\usepackage{bm,nicefrac}

\usepackage[T1]{fontenc}
\usepackage{ae,aecompl}

\usepackage{newtxtext,newtxmath}

\title[Ground-Based Transmission Spectroscopy of WASP-103b]{\textit{Ground-Based Transmission Spectroscopy with FORS2}: \\ A featureless optical transmission spectrum and detection of H$_2$O for the ultra-hot Jupiter WASP-103b}

\author[J. Wilson et al.]{Jamie Wilson$^{1}$\thanks{Contact e-mail: \href{mailto:jwilson34@qub.ac.uk}{jwilson34@qub.ac.uk}}, {Neale P. Gibson$^{2}$}, {Nikolay Nikolov$^{3}$}, {Savvas Constantinou$^{4}$},\newauthor
{Nikku Madhusudhan$^{4}$}, {Jayesh Goyal$^{5}$}, {Joanna K. Barstow$^{6}$}, {Aarynn L. Carter$^{7}$}, \newauthor
{Ernst J. W. de Mooij$^{1}$}, {Benjamin Drummond$^{7}$}, {Thomas
 Mikal-Evans$^{8}$}, \newauthor {Christiane Helling$^{9}$},
 {Nathan J. Mayne$^{7}$}
 and {David K. Sing$^{10}$}
\smallskip
\\
$^{1}$Astrophysics Research Centre, School of Mathematics and Physics, Queens University Belfast, Belfast BT7 1NN, UK\\
$^{2}$School of Physics,  Trinity College Dublin, The University of Dublin, Dublin 2, Ireland\\
$^{3}$Space Telescope Science Institute, 3700 San Martin Drive, Baltimore, MD 21218, USA\\ $^{4}$Institute of Astronomy, University of Cambridge, Madingley Road, Cambridge CB3 0HA, UK\\ $^{5}$Carl Sagan Institute, Department of Astronomy, Cornell University, Ithaca, NY 14850, USA\\
$^{6}$The Open University, Walton Hall, Kents Hill, Milton Keynes, MK7 6AA, UK\\
$^{7}$Astrophysics Group, University of Exeter, Physics Building, Stocker Road, Devon, EX4 4QL, UK\\
$^{8}$Kavli Institute for Astrophysics
 and Space Research, Massachusetts Institute of Technology, 77 Massachusetts Avenue, 37-241, Cambridge, MA 02139, USA\\
$^{9}$Centre for Exoplanet Science, University of St Andrews, Nort Haugh, St Andrews, KY169SS, UK\\
$^{10}$Department of Earth and Planetary Sciences, Johns Hopkins University, Baltimore, MD 21218, USA}

\date{Last updated 2015 May 22; in original form 2013 September 5}

\pubyear{2020}

\begin{document}
\label{firstpage}
\pagerange{\pageref{firstpage}--\pageref{lastpage}}
\maketitle

\begin{abstract}
We report ground-based transmission spectroscopy of the highly irradiated and ultra-short period hot-Jupiter WASP-103b covering the wavelength range $\approx$\,400\,--\,600\,nm using the FORS2 instrument on the Very Large Telescope. The light curves show significant time-correlated noise which is mainly invariant in wavelength and which we model using a Gaussian process. The precision of our transmission spectrum is improved by applying a common-mode correction derived from the white light curve, reaching typical uncertainties in transit depth of $\approx$\,$2\times10^{-4}$ in wavelength bins of 15\,nm. After correction for flux contamination from a blended companion star, our observations reveal a featureless spectrum across the full range of the FORS2 observations and we are unable to confirm the Na absorption previously inferred using Gemini/GMOS or the strong Rayleigh scattering observed using broad-band light curves. We performed a Bayesian atmospheric retrieval on the full optical-infrared transmission spectrum using the additional data from Gemini/GMOS, HST/WFC3 and Spitzer observations and recover evidence for H$_2$O absorption at the 4.0\,$\sigma$ level. However, our observations are not able to completely rule out the presence of Na, which is found at 2.0\,$\sigma$ in our retrievals. This may in part be explained by patchy/inhomogeneous clouds or hazes damping any absorption features in our FORS2 spectrum, but an inherently small scale height also makes this feature challenging to probe from the ground. Our results nonetheless demonstrate the continuing potential of ground-based observations for investigating exoplanet atmospheres and emphasise the need for the application of consistent and robust statistical techniques to low-resolution spectra in the presence of instrumental systematics.

\end{abstract}




\begingroup
\let\clearpage\relax

\endgroup
\newpage

\section{Introduction}

Transmission spectroscopy has proven to be a highly successful method for probing the atmospheres of close-in exoplanets, allowing us to infer the chemical composition and physical structure of a planet's atmosphere without needing to spatially resolve the planet and star. During primary transit, when a planet crosses the disk of a star from the point of view of an observer, a small fraction of the stellar light is filtered through the annulus of the planet's atmosphere \citep{2001ApJ...553.1006B,2000ApJ...537..916S} with the observed transit depth increasing at wavelengths corresponding to strong atomic and molecular absorption. The transit depth as a function of wavelength (conventionally measured as a planet-to-star radius ratio) is known as a transmission spectrum and is sensitive to compositions along the day-night terminator of the planet. 

Transit and radial velocity surveys have revealed that a significant subset of exoplanetary systems are surprisingly unlike anything found in our own Solar System, and show a remarkably diverse range of properties. This includes the discovery of highly irradiated hot-Jupiters \citep[e.g.][]{1995Natur.378..355M,2000ApJ...529L..41H,2000ApJ...529L..45C} - gas giants with masses similar to Jupiter orbiting extraordinarily close to their host stars - which exhibit a wide variety of transmission spectra and a continuum from clear to cloudy atmospheres \citep{2016Natur.529...59S}.

Substantial progress in the area of exoplanet atmospheric characterisation was first achieved using space-based instruments such as the \textit{Hubble Space Telescope} (HST) \citep[e.g.][]{2002ApJ...568..377C,2008MNRAS.385..109P,2012MNRAS.422.2477H,2012ApJ...747...35B,2013MNRAS.432.2917P,2014Natur.505...69K,2015MNRAS.447..463N} and \textit{Spitzer Space Telescope} \citep[e.g.][]{2005ApJ...626..523C,2005Natur.434..740D,2007Natur.447..183K,2007ApJ...661L.191B,2013ApJ...776L..25D}, but ground-based observations, utilising multi-object differential spectrophotometry, have been rapidly catching up with their own significant contributions \citep[e.g.][]{Snellen_2008,2008ApJ...673L..87R,2010Natur.468..669B,2013A&A...559A..33C,2013MNRAS.436.2974G,2013MNRAS.428.3680G,2016MNRAS.463.2922K,2016A&A...587A..67L,2016A&A...590A.100M,2014Sci...346..838S}. The importance of ground-based observations for exoplanetary science is set to continue well into the era of the upcoming James Webb Space Telescope (JWST) by complementing the newly acquired near- and mid-IR observations with those obtained in the optical regime.

Here we report ground-based transmission spectroscopy results for the ultra-hot Jupiter WASP-103b using the FOcal Reducer and Spectrograph (FORS2) mounted on the European Southern Observatory's (ESO) Very Large Telescope (VLT). FORS2 is a general-purpose imager, spectrograph and polarimeter \citep{1998Msngr..94....1A} which has been shown to offer improved performance for exoplanet spectroscopy after undergoing an upgrade to its Linear Atmospheric Dispersion Corrector \citep{2016SPIE.9908E..2BB}, with detections of Na and K absorption and scattering by clouds and hazes in multiple exoplanet atmospheres  \citep[e.g.][]{2015A&A...576L..11S,2016ApJ...832..191N,2018Natur.557..526N}. Our results are part of a large, ground-based, comparative survey which aims to study the chemical compositions and occurrence rates of clouds and hazes over the full range of mass and temperature regimes \citep[e.g.][]{2016ApJ...832..191N,2017MNRAS.467.4591G,2020MNRAS.tmp.1223C}.

WASP-103b is an ultra-short period (P = 0.9\,d), highly irradiated (\textit{T}$_\mathrm{eq}$\,$\approx$\,2500\,K) hot-Jupiter discovered by \citet{2014A&A...562L...3G}. It has a mass and radius significantly larger than Jupiter - 1.49\,\textit{M}$_\mathrm{J}$ and 1.53\,\textit{R}$_\mathrm{J}$ respectively - and transits a late F-type (V\,$\approx$\,12.1) main-sequence star. At a separation of less than 1.2 times the Roche limit WASP-103b is expected to be in the late stages of orbital decay and close to tidal disruption \citep[e.g.][]{Matsumura_2010,2017AJ....154....4P}. \citet{Staab_2016} measured the chromospheric activity of WASP-103 finding marginal evidence that it was higher than expected from the system age (log(\textit{R$'$}$_\mathrm{HK}$) = -4.57). \citet{Pass_2019} found a dayside effective temperature of $\approx$\,3200\,K using Gaussian process regression on WFC3 and Spitzer secondary eclipse depth measurements. Meanwhile, \citet{Garhart_2020} calculated an effective temperature of $\approx$\,2500\,K and measured brightness temperatures in the 3.6 micron and 4.5 micron Spitzer bands of $\approx$\,2800\,K and $\approx$\,3100\,K respectively. 

Follow-up observations by \citet{2015MNRAS.447..711S} revealed a strong wavelength-dependent slope in their broad-band optical transmission spectrum which they concluded was too steep to be caused by Rayleigh scattering processes alone. A re-analysis of the same data by \citet{2016MNRAS.463...37S} accounting for the flux contamination of a previously unknown companion star \citep{2015A&A...579A.129W} instead showed a minimum around 760\,nm and increasing opacity towards both the blue and red. This overall picture was subsequently confirmed by \citet{2018MNRAS.474.2334D} from an independent global analysis including a large fraction of the same archival transit light curves. This surprising V-shaped transmission spectrum cannot be easily explained by theoretical models and nor is it confirmed by higher-resolution observations with Gemini/GMOS, which instead showed signs of enhanced absorption in the cores of the Na and K features \citep{2017A&A...606A..18L} and no evidence for a Rayleigh scattering signature, suggesting that WASP-103b might possess a largely clear atmosphere at the terminator region. However, since they did not have any data bluewards of 550\,nm they were unable to conclusively rule out the presence of a scattering slope. 

In the near-IR, \citet{2017AJ....153...34C} found a featureless emission spectrum using HST/WFC3 which was indistinguishable from that due to an isothermal atmosphere and could be explained by either a thermal inversion layer or clouds and/or hazes in the upper atmosphere and suggested the need for additional optical observations in order to differentiate between these possible explanations. \citet{2018AJ....156...17K} observed a featureless transmission spectrum between 1.15 and 1.65\,$\mu$m with WFC3/Spitzer at the 1\,$\sigma$ level after correcting for nightside emission and determined that their phase-resolved spectra were consistent with blackbody emission at all orbital phases, attributing the lack of detection of dayside spectral features of water to partial H$_2$O dissociation.

This paper is structured as follows: we describe our observations and data reduction steps in Section 2 and detail our light curve analysis and contaminant correction in Section 3; in Section 4 we describe our atmospheric modelling approach and discuss our results in Section 5. Finally, we offer our conclusions in Section 6.

\section{FORS2 Observations and Data Reduction}
We observed a single transit of the hot-Jupiter WASP-103b during the night of 2017 May 1st with the FORS2 spectrograph mounted on the 8.2\,--\,m `Antu' telescope of the VLT at the European Southern Observatory, Paranal, Chile, as part of the large program 199.C-0467 (PI: Nikolov). Our transit was observed using the GRIS600B (hereafter 600B) grating covering the spectral range of 320\,--\,620\,nm with a total of 174 science exposures of 80\,s each, covering a total period of 310 minutes with a readout time of $\sim$\,30\,s. FORS2 consists of two 2k\,$\times$\,4k CCDs separated by a small detector gap with an image scale of 0.25$''$\,/\,pixel in 2\,$\times$\,2 binning mode, corresponding to a field-of-view of 6.8\,$\times$\,6.8 arcminutes squared.

Observations of the target and two comparison stars were carried out simultaneously in multi-object (MXU) spectroscopy mode. We used a custom mask consisting of broad slits accurately centred on the positions of WASP-103 and the comparison stars with a width of 22$''$ and length of 120$''$ to reduce differential slit losses from seeing variations and guiding inaccuracies. We found that one of our comparison stars was significantly fainter than the other and so we excluded this from our analysis and only used the brighter of the two stars. FWHM for the observations was typically $\sim$\,3 pixels but reached a maximum of $\sim$\,7 pixels towards the very beginning of the observations resulting in seeing-limited resolution of R\,$\approx$\,450\,--\,1050, with airmass varying from a maximum of 1.87 at the commencement of observations down to 1.18. 

We used the FORS2 pipeline for standard bias and flat field corrections with relevant calibration frames taken before and after the science exposures. However, we found that neither of these corrections had a significant influence on our conclusions and therefore we proceeded using only the raw frames for our final analysis. Spectral extraction was performed in IRAF\footnote{IRAF is distributed by the National Optical Astronomy Observatory, which is operated by the Association of Universities for Research in Astronomy (AURA) under cooperative agreement with the National Science Foundation}/PyRAF\footnote{PyRAF is a product of the Space Telescope Science Institute, which is operated by AURA for NASA} using a custom pipeline and summing an aperture radius of 15 pixels after background subtraction (we found that a radius of 15 pixels resulted in the lowest average uncertainties for our transmission spectrum). We estimated the background contribution by taking the median value in a region of pixels located 80\,--\,100 pixels either side of the spectral trace. Example spectra of WASP-103 and the reference star are shown in Figure 1.

Wavelength calibration was performed using arc lamp exposures with a calibration mask in place which is identical to the science mask but with narrower 1$''$ slit widths to obtain arcs with narrower features for more precise calibration. We accounted for shifts in the dispersion direction by cross-correlating the target spectra using the H$\beta$ line after normalising the continua, and then cross-correlating again between the target and comparison star using the same feature. We then used the measured x-shifts to realign all spectra to the reference spectrum's wavelength scale. To check that our results were not overly sensitive to the specific choice of feature we also tried extracting the x-shifts by cross-correlating using the Na feature, but found that this had little impact on our final transmission spectrum, and therefore we present our results using only the H$\beta$ alignment. 

We found that the wavelength solution obtained from the reduction pipeline resulted in small residual offsets between our target and comparison star and so we decided to construct an alternative solution using a set of well-resolved lines in the mean spectrum (after realignment) and fitting gaussians to each of these lines to accurately determine the line centres. We used a Gaussian process (GP) to fit the measured line centres. GPs are routinely used within the machine learning community for Bayesian non-parametric regression problems and were introduced by \citet{2012MNRAS.419.2683G} for the analysis of systematics in exoplanet time-series. We discuss our implementation of GPs in Section 3.1. We also tried fitting using a second-order polynomial but obtained near identical results. In principle we could fit with a higher order polynomial but this is unlikely to alter our final transmission spectrum given that the changes are small when compared to our bin widths and we proceeded using the wavelength solution derived from the GP fit. 

\begin{figure}
 \includegraphics[width=\columnwidth]{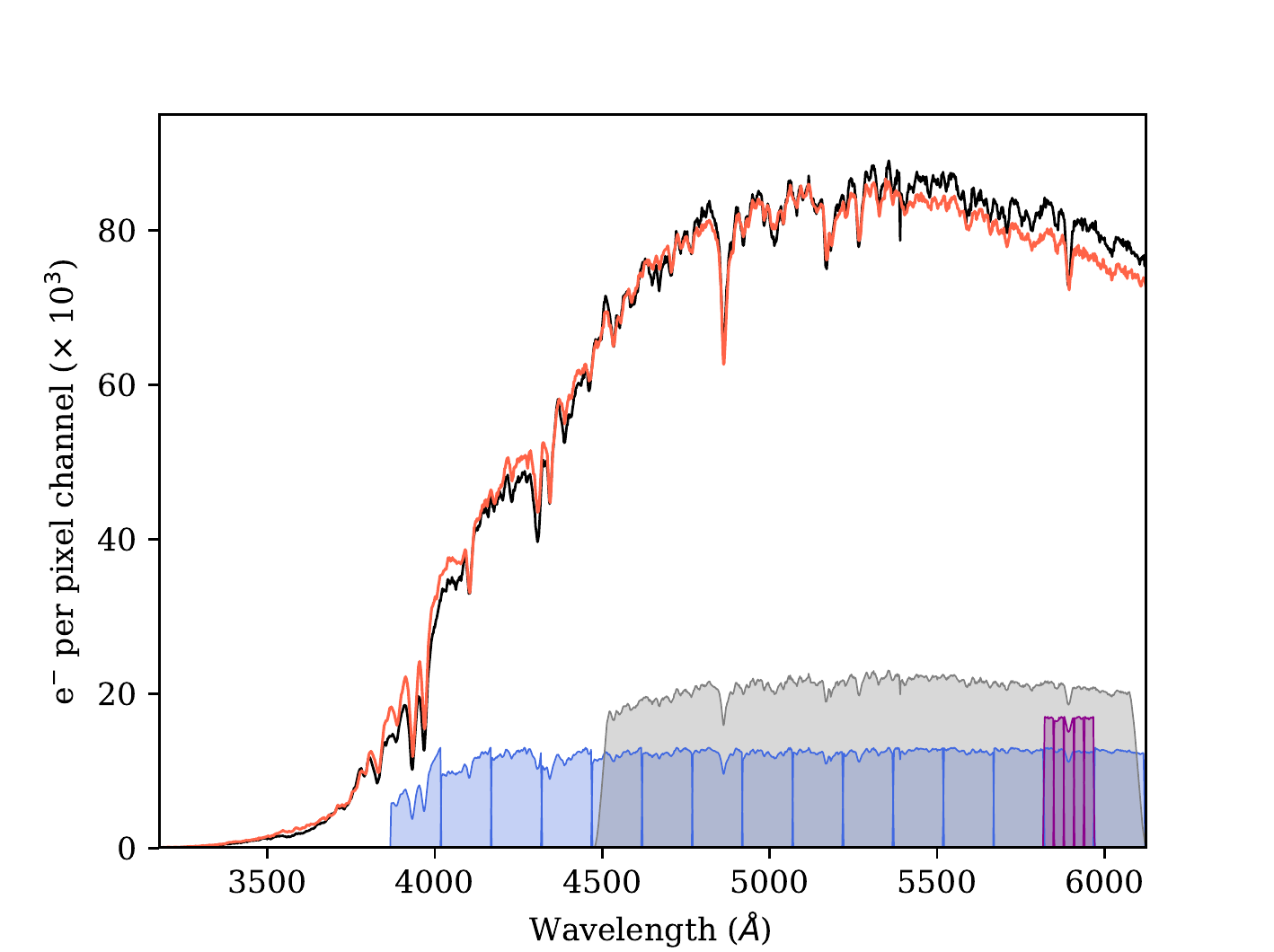}
 \caption{Example spectra of the target (black) and one reference star (red). Coloured regions indicate the spectral bins used for extraction of the white light curve (grey), spectroscopic light curves (blue) and the high-resolution bins centred around the Na feature (magenta).}
 \label{fig:examplespec}
\end{figure}

The time-series spectra were then used to construct the white light curve by summing the flux of each stellar spectrum over a broad wavelength range as shown in Figure 1, and dividing the target star's flux by the comparison star's flux, thereby correcting for the effects of atmospheric transparency variations. We also constructed multiple `spectral' light curves by integrating over the narrower channels also shown in Figure 1 and discussed in Section 3.2. We tried varying the total number of channels used in our analysis and found that, while this altered the resolution and signal-to-noise of our results, it did not significantly influence our conclusions. In the end we chose to extract a total of 15 individual wavelength channels and the resulting white light curve and spectral light curves are shown in Figures 2 and 3.

We also calculated the theoretical noise for our white light curve and each of our spectral light curves, including the contributions from photon noise, read noise and the sky background. The average electron counts per exposure for the white light curve was $\approx$\,9\,$\times$\,10$^{7}$ for both the target and comparison star resulting in time-averaged theoretical precision in the relative flux per exposure of $\approx$\,1.4\,$\times\,$10$^{-4}$. The average electron counts per exposure for the spectral light curves varied from $\approx$\,2\,$\times\,$10$^{6}$ to $\approx$\,1\,$\times\,$10$^{7}$ for the target star and $\approx$\,3\,$\times\,$10$^{6}$ to $\approx$\,9\,$\times\,$10$^{6}$ for the comparison. The time-averaged theoretical precision in the relative flux per exposure for the spectral light curves therefore ranges from $\approx$\,4\,$\times$\,10$^{-4}$ to $\approx$\,9\,$\times$\,10$^{-4}$.

Finally, we also extracted auxiliary measurements from the target and comparison spectra, including the shifts in the dispersion and cross-dispersion axes and the width of the spectral trace. Such measurements can in principle be used to attempt to investigate the cause of the instrument systematics in the light curves \citep[e.g.][]{2001ApJ...553.1006B,2003nicm.rept....1G,2007A&A...476.1347P,2008arXiv0812.1844S,2010Natur.464.1161S,2012A&A...542A...4G,2013MNRAS.434.3252H,Nikolov_2016}, however in our case we found no obvious correlations between the auxiliary measurements and the form of the systematics.

We used the PyLDTk toolkit \citep{2015MNRAS.453.3821P}, which uses the spectral libraries of \citet{2013A&A...553A...6H}, to determine the limb darkening parameters for the spectral response functions (adopting the stellar values for WASP-103), and used the system parameters and uncertainties for WASP-103b given in the discovery paper \citep{2014A&A...562L...3G}.

\section{Analysis}
\subsection{White Light Curve Analysis}

Rather than impose a prespecified parametric form to describe the unknown instrumental systematics, we follow the procedure described by \citet{2012MNRAS.419.2683G} and use a time-dependent GP\footnote{For the implementation of our Bayesian inference we made extensive use of the Python modules \textbf{GeaPea} and \textbf{Infer} which are freely available from \url{https://github.com/nealegibson}} to model the systematics as a stochastic process simultaneously with a deterministic transit model derived from the equations of \citet{2002ApJ...580L.171M}. This approach leads to a much more flexible model for the instrumental effects, and being intrinsically Bayesian, automatically helps mitigate against the possibility of over-fitting. In our case a GP defines a joint Gaussian probability distribution around a transit mean function given by:
\begin{equation}
p(\bmath f| \bmath t,\bphi,\btheta) = \mathcal{N} \left (T(\bmath t,\bphi) , \boldsymbol{\Sigma} (\bmath t,\btheta) \right).
\end{equation}
where $\bmath t$ is the vector of time measurements, \textit{T} is the transit function depending on $\bmath t$ and the transit parameters $\bphi$, $\bmath f$ is the vector of flux measurements and $\boldsymbol{\Sigma}$ is the covariance matrix which is a function of $\bmath t$ and the hyperparameters $\btheta$. GPs are capable of including multiple inputs such as the optical state parameters which describe the behaviour of the instrument, however given that we found no obvious correlations between the instrumental systematics and auxiliary measurements we proceeded to model the systematics as time-correlated noise only. The instrumental systematics are fully described by the covariance matrix which describes the correlation between data points, and the covariance matrix itself is populated by the covariance function, also known as a kernel (see \citet{3569} for a detailed discussion of kernels), with parameters $\boldsymbol{\theta}$. For our analysis we used the Mat\'ern 3/2 kernel defined as:
\begin{equation}
    k({t}_n, {t}_m | \btheta) = \xi^2 \left( 1+{\sqrt{3}\,\eta\,\Delta t} \right) \exp \left( -{\sqrt{3}\,\eta\,\Delta t}\right) + \delta_{nm}\sigma^2,
\end{equation}
where $\xi$ specifies the maximum covariance or height scale, $\Delta$\,t is the time difference of observations, $\eta$ is the inverse characteristic length scale, $\delta_{nm}$ is the Kronecker delta and $\sigma$ specifies the white noise (assumed to be identical for all data points). The Mat\'ern 3/2 kernel can be viewed as a less smooth version of the more commonly employed squared exponential kernel and our choice was mainly motivated by the arguments outlined in \citet{2013MNRAS.436.2974G}. As a check we also ran the same analysis using the squared exponential kernel but found that this had little effect on the final results. The posterior probability distribution is then obtained by specifying priors for the hyperparameters of the model and multiplying by the marginal likelihood (in practice we use log priors and the log marginal likelihood).

Our mean function is the deterministic transit model assuming a circular orbit and the two parameter quadratic limb darkening law of \citet{2000A&A...363.1081C} with coefficients \textit{c}$_\mathrm{1}$ and \textit{c}$_\mathrm{2}$. In our analysis we held the value of the period fixed to that reported by \citet{2014A&A...562L...3G} and fit for the central transit time (\textit{T}$_\mathrm{c}$), planet-to-star radius ratio ($\rho$ = \textit{R}$_\mathrm{p}$/\textit{R}$_\mathrm{\star}$) and a further two parameters of a linear baseline model of time (\textit{f}$_{\mathrm{oot}}$, \textit{T}$_{\mathrm{grad}}$). We chose to fix the values for the system scale (\textit{a}/\textit{R}$_\mathrm{\star}$) and the impact parameter \textit{b} at the tightly constrained values reported in \citet{2015MNRAS.447..711S} and set a Gaussian prior for the planet-to-star radius ratio also using their reported value. This was to help facilitate a direct comparison with the results from \citet{2018AJ....156...17K} who adopted these parameter values for their analysis. This constrains the white light curve parameters to previously derived values and enables us to recover a more accurate systematics model. As a test, we also performed an independent fit to our white light curve to check the validity of our assumed parameter values. Both of the parameters which we chose to fix in our analysis were found to be consistent within 1\,$\sigma$ to those of \citet{2015MNRAS.447..711S}, though the measured planet-to-star radius ratio was found to be slightly higher (within 2\,$\sigma$). However, since our retrieval accounts for an offset in the planet-to-star radius ratio between the different datasets (see Section 4.2 for a description of our atmospheric retrieval using AURA), we don't expect this discrepancy to significantly affect our results. In addition, common-mode corrections can lead to biases in the mean level of the transmission spectrum for each instrument/transit if the correction is inaccurate. This does not affect the relative transmission spectrum for each individual transit observation, but can lead to offsets between datasets which should be taken into account in the interpretation.

We also placed Gaussian priors on the limb darkening parameters \textit{c}$_\mathrm{1}$ and \textit{c}$_\mathrm{2}$ with a mean and uncertainty determined from the best fit values from PyLDTk, and additionally restricted their values to ensure that the brightness of the stellar surface is positive with a monotonically decreasing intensity profile using the following boundary conditions \citep[e.g.][]{Kipping_2013}:
\begin{eqnarray}
    \textit{c}_\mathrm{1}+\textit{c}_\mathrm{2} < 1,\nonumber \\
    \textit{c}_\mathrm{1} > 0,\nonumber \\
    \textit{c}_\mathrm{1}+2\textit{c}_\mathrm{2} > 0.
\end{eqnarray}
As another check we also repeated our analysis having fixed the limb darkening parameters to their best fit values but found this did not affect the conclusions of our study. We summarise the assumed values for the white light curve in Table 1. The kernel hyperparameters are variable in our fit but we fit for log $\xi$ and log $\eta$ with uniform priors in log space which is the natural parameterisation for scale parameters \citep[e.g.][]{2013MNRAS.428.3680G,2013MNRAS.436.2974G}. We also constrain the length scale to be no smaller than the cadence of our observations and no larger than twice the total duration with lower frequency systematics being accounted for in the baseline function.

\begin{figure}
 \includegraphics[width=\columnwidth]{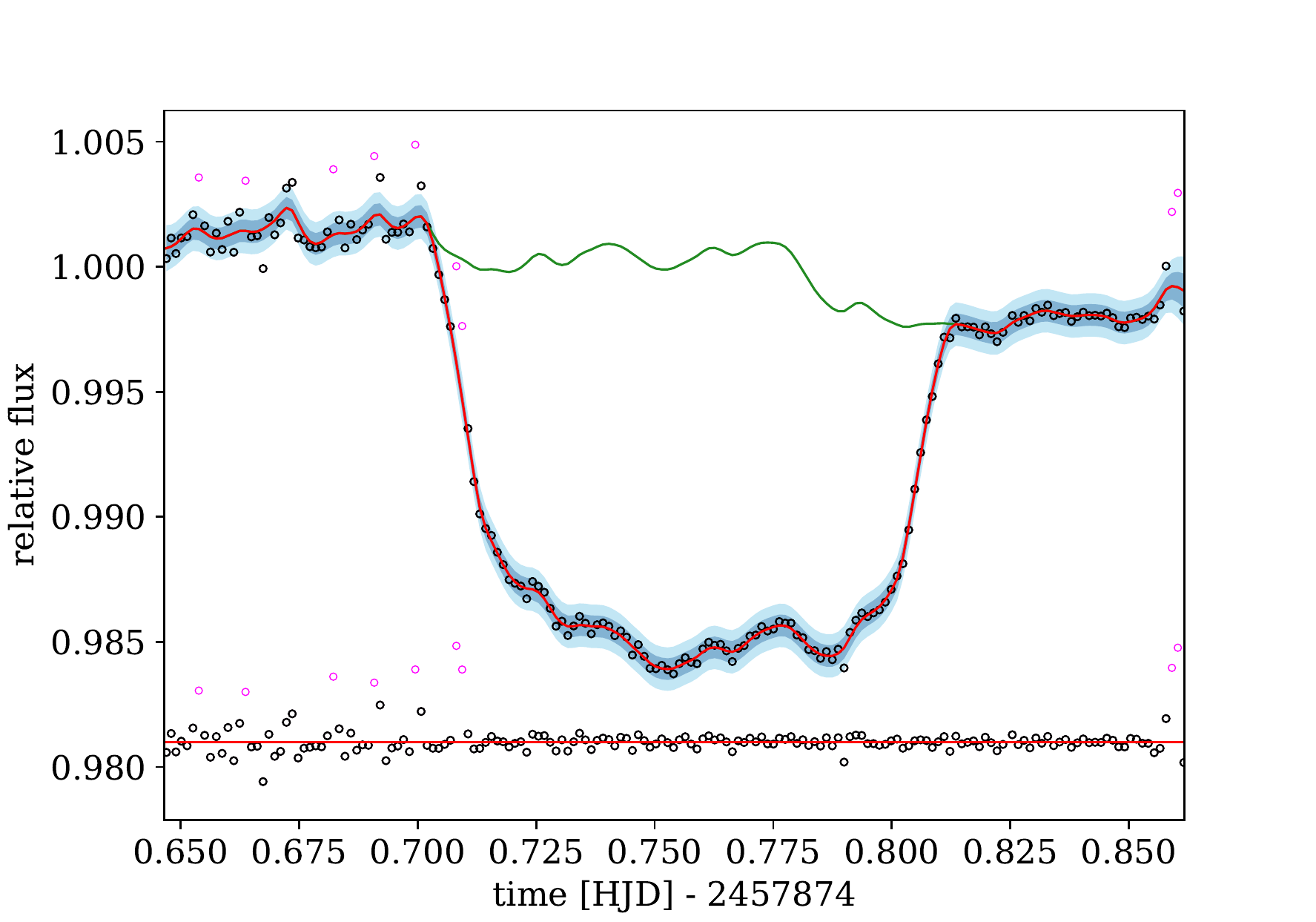}
 \caption{White light curve of WASP-103b obtained with the 600B grism. The red line shows the best fit model with blue shading indicating plus/minus two standard deviations. The green line shows the systematics model derived from the GP fit. Residuals are indicated below the light curve. We clip any points over 4\,$\sigma$ from the fit, but preserve them for the common-mode correction (shown in magenta, see Section 3.1). }
 \label{fig:wlc}
\end{figure}

Our best-fitting model for the white light curve is obtained by optimising the posterior over the transit and kernel parameters using a differential evolution algorithm with the values from \citet{2014A&A...562L...3G} and \citet{2015MNRAS.447..711S} as the starting point, and then fine-tuning our estimated values using a Nelder-Mead simplex algorithm. Due to some high-frequency systematics (most likely caused by thin clouds), which occur before and at the beginning of ingress, we clip any data points over 4\,$\sigma$ from our white light curve fits, to avoid biasing our systematics model towards short length scales. Nonetheless, the clipped points are retained in the common-mode correction, to correct similar high-frequency systematics also present in the spectroscopic light curves. We verified that this process did not significantly affect our final transmission spectrum by obtaining near identical results when the points are included in the GP fit. We then marginalise our posterior distribution using a Markov-Chain Monte-Carlo (MCMC) method to obtain uncertainty estimates for our parameters. For each of our light curves we used 4 independent chains of length 80,000, discarding the first 40$\%$ of samples in the chain and checking for mutual convergence using the Gelman-Rubin statistic. We derive our best-fit systematics model by separating the mean of our GP (conditioned on the observed data) from the transit model and use this for our common-mode correction for the spectroscopic light curves. The best fit white light curve model and derived systematics model and residuals are shown in Figure 2. 

\begin{table}
 \caption{Transit parameter values used in the fitting of the white light curve. The orbital period, system scale and impact parameter were held fixed and Gaussian priors were placed on the following parameters with the mean and standard deviations given below.}
 \label{tab1}
 \begin{tabular}{ll}
  \hline
  Parameter & Value\\
  \hline
  \textit{P} & 0.925542 days (fixed)\\[2pt] 
  \textit{a}/\textit{R}$_\mathrm{\star}$ & 2.999 (fixed)\\[2pt]
  \textit{b} & 0.14 (fixed)\\[2pt]
  \textit{$\rho$} & 0.1127 $\pm$ 0.0009\\[2pt]
  \textit{c}$_\mathrm{1}$ & 0.614 $\pm$ 0.004\\[2pt]
  \textit{c}$_\mathrm{2}$ & 0.102 $\pm$ 0.005\\[2pt]
  \hline
 \end{tabular}
\end{table}

\subsection{Spectroscopic Light Curve Analysis}

\begin{figure*}
 \includegraphics[width=\textwidth]{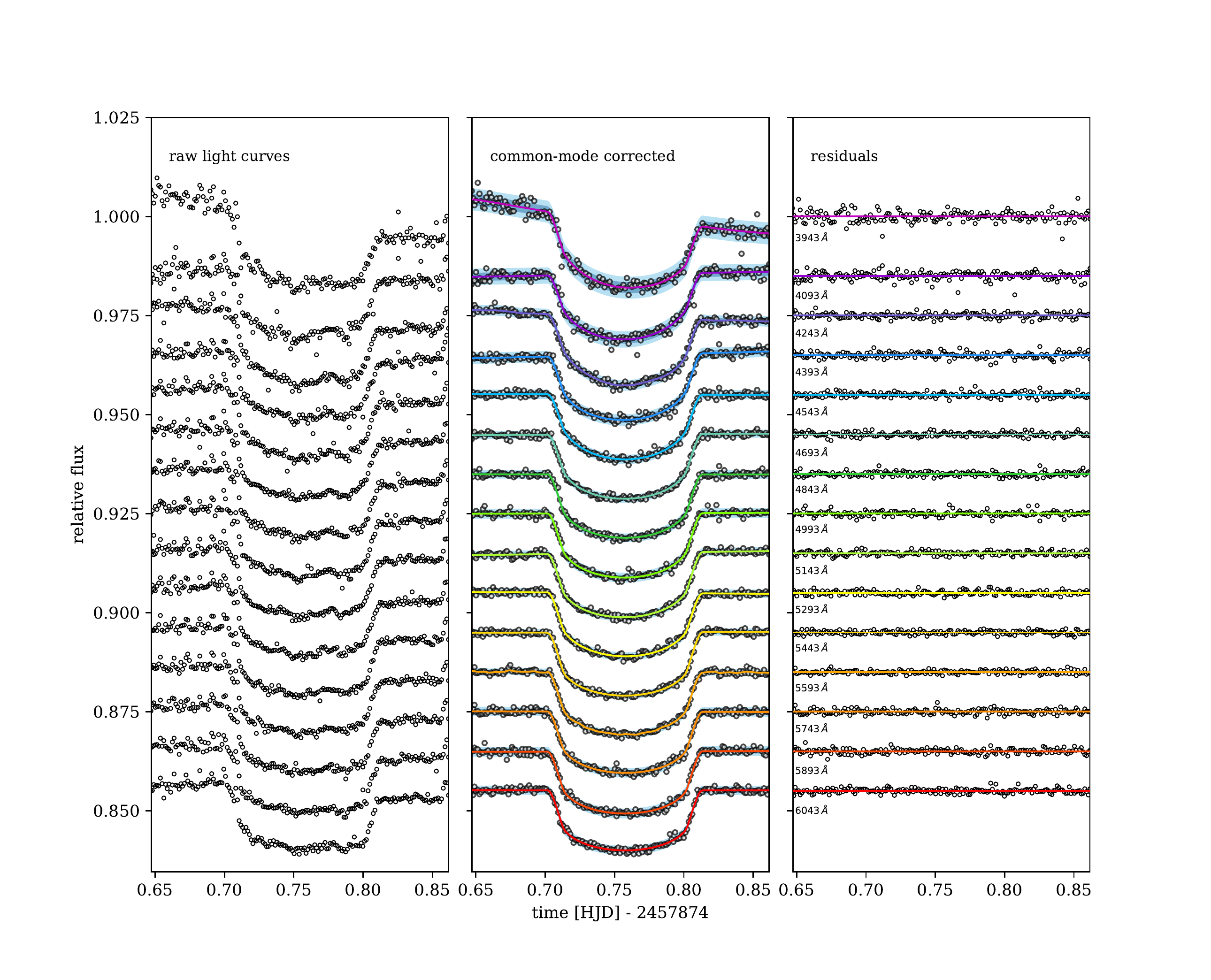}
 \caption{Spectral light curves for the 600B grism corresponding to the broad spectral channels shown in Figure 1. The left panel shows the raw light curves before correction. The middle panel shows the light curves with best fit GP model after the common-mode correction. The right panel shows the residuals from the best-fit model.}
 \label{fig:slc1}
\end{figure*}

For the spectroscopic light curves, we first extracted individual low-resolution channels using uniform bins with a width of 150\,{\AA} as shown in Figure 1. In total we extracted 15 of these low-resolution channels and the resulting light curves are shown in the left panel of Figure 3. The spectroscopic light curves are corrupted by significant systematics which are similar in shape to that seen for the white light curve, and which are mainly invariant in wavelength. This allows us to correct the spectral light curves prior to model fitting by dividing through by the common-mode correction which we derive from the white light curve. We also subtract the residuals from the white light curve and its best-fitting model to remove any remaining high-frequency systematics. This process removes much of the common signal in the light curves and results in significant improvements to the precision of our transmission spectrum without affecting the relative value of the planet-to-star radius ratio.

After correction we fit each spectroscopic light curve using the same method outlined above for the white light curve except we fix the central transit time to the best fit value inferred from the white light curve analysis, and allow the planet-to-star radius ratio, limb darkening parameters, normalisation parameters and kernel hyperparameters to vary for each light curve fit. We set broad normal priors for the limb darkening coefficients centred at the best fit values determined using PyLDTk, but increased the uncertainties for our prior to have a standard deviation of 0.1 in order to allow a greater degree of flexibility in our model. We again optimise the transit and kernel parameters using a differential evolution algorithm and fine-tuned using a Nelder-Mead simplex algorithm. We clip any outliers over 4\,$\sigma$ from our predictive distribution for each individual fit (typically only 1-2 points for each light curve) before running the same MCMC procedure to explore our posterior distribution as for the white light curve.

The best fit GP models are shown in Figure 3 and we summarise the derived planet-to-star radius ratios and associated uncertainties in Tables 2 and 3.

\begin{table}
 \caption{Transmission spectrum for WASP-103b recovered from the FORS2 low-resolution spectroscopic light curves. }
 \label{tab2}
 \begin{tabular}{lcc}
  \hline
  Wavelength  & Radius Ratio & Limb Darkening\\
  Centre [Range] ({\AA}) & \textit{R}$_\mathrm{p}$/\textit{R}$_\mathrm{\star}$ & \textit{c}$_\mathrm{1}$ \hspace{0.5cm} \textit{c}$_\mathrm{2}$\\
  \hline
  3943 [3868-4018] & 0.11352 $\pm$ 0.00292 & 0.860\hspace{0.3cm}-0.066\\
  4093 [4018-4168] & 0.11076 $\pm$ 0.00100 & 0.797\hspace{0.3cm}0.020\\
  4243 [4168-4318] & 0.11241 $\pm$ 0.00140 & 0.856\hspace{0.3cm}-0.057 \\
  4393 [4318-4468] & 0.11162 $\pm$ 0.00105 & 0.749\hspace{0.3cm}0.035\\
  4543 [4468-4618] & 0.11095 $\pm$ 0.00068 & 0.730\hspace{0.3cm}0.053\\
  4693 [4618-4768] & 0.11205 $\pm$ 0.00047 & 0.696\hspace{0.3cm}0.078\\
  4843 [4768-4918] & 0.11168 $\pm$ 0.00085 & 0.631\hspace{0.3cm}0.111\\
  4993 [4918-5068] & 0.11280 $\pm$ 0.00059 & 0.645\hspace{0.3cm}0.093\\
  5143 [5068-5218] & 0.11309 $\pm$ 0.00049 & 0.624\hspace{0.3cm}0.094\\
  5293 [5218-5368] & 0.11176 $\pm$ 0.00062 & 0.600\hspace{0.3cm}0.107\\
  5443 [5368-5518] & 0.11330 $\pm$ 0.00045 & 0.581\hspace{0.3cm}0.109\\
  5593 [5518-5668] & 0.11143 $\pm$ 0.00057 & 0.563\hspace{0.3cm}0.119\\
  5743 [5668-5818] & 0.11103 $\pm$ 0.00055 & 0.544\hspace{0.3cm}0.125\\
  5893 [5818-5968] & 0.11163 $\pm$ 0.00101 & 0.529\hspace{0.3cm}0.128 \\
  6043 [5968-6118] & 0.11106 $\pm$ 0.00092 & 0.516\hspace{0.3cm}0.129\\
  \hline
 \end{tabular}
\end{table}

\begin{table}
 \caption{Transmission spectrum for WASP-103b recovered from the FORS2 high-resolution spectroscopic light curves centered on the Na feature. }
 \label{tab3}
 \begin{tabular}{lcc}
  \hline
  Wavelength  & Radius Ratio & Limb Darkening\\
  Centre [Range] ({\AA}) & \textit{R}$_\mathrm{p}$/\textit{R}$_\mathrm{\star}$ & \textit{c}$_\mathrm{1}$\ \hspace{0.5cm} \textit{c}$_\mathrm{2}$\\
  \hline
  5833 [5818-5848] & 0.11154 $\pm$ 0.00117 & 0.535\hspace{0.3cm}0.130\\
  5863 [5848-5878] & 0.11240 $\pm$ 0.00105 & 0.525\hspace{0.3cm}0.131\\
  5893 [5878-5908] & 0.11329 $\pm$ 0.00137 & 0.535\hspace{0.3cm}0.121\\
  5923 [5908-5938] & 0.11187 $\pm$ 0.00117 & 0.528\hspace{0.3cm}0.130\\
  5953 [5938-5968] & 0.11112 $\pm$ 0.00177 & 0.524\hspace{0.3cm}0.129\\

  \hline
 \end{tabular}
\end{table}

\subsection{Investigation of sodium feature}

\begin{figure*}
 \includegraphics[width=\textwidth]{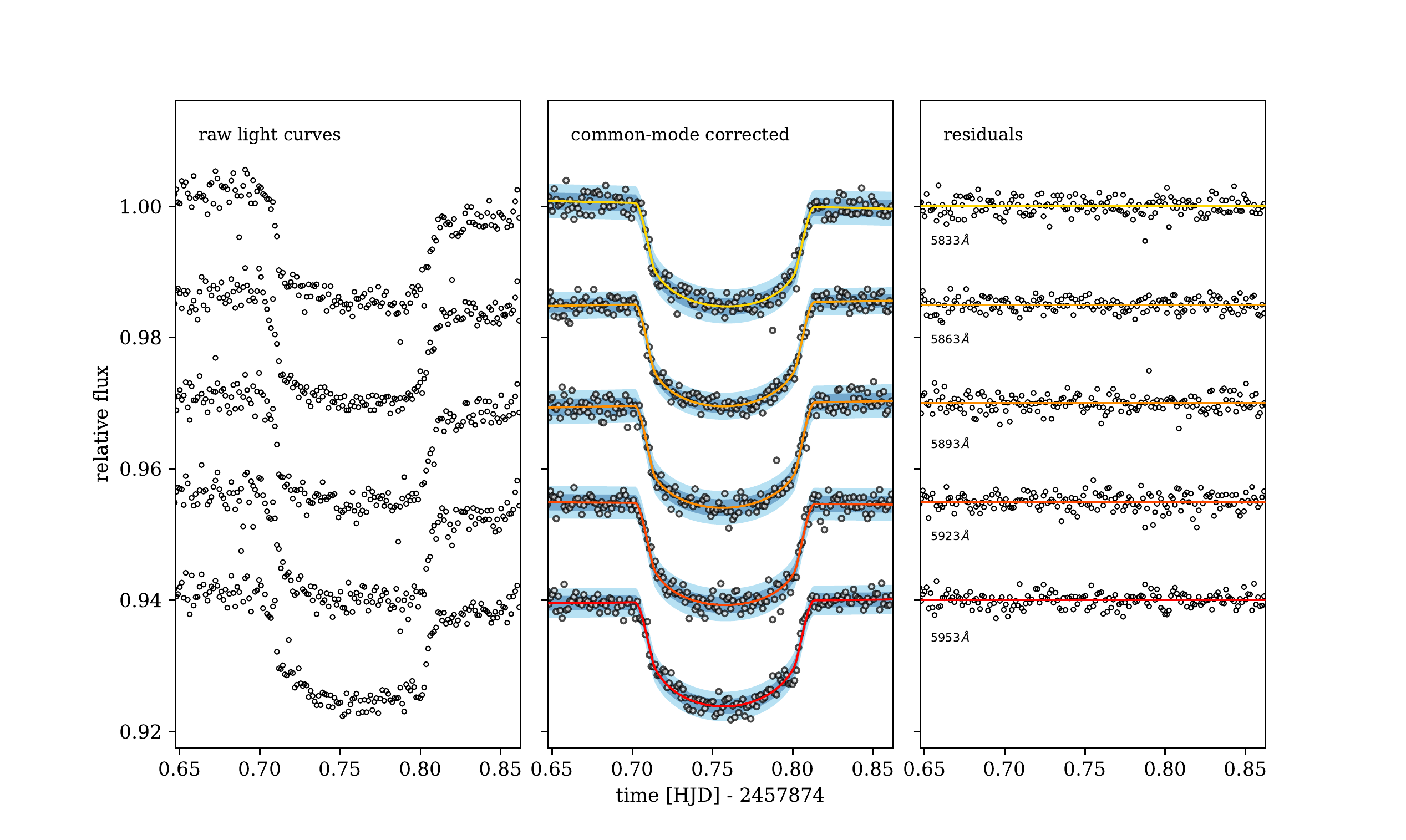}
 \caption{Same as Figure 3, showing the five additional light curves extracted using the high-resolution channels shown in Figure 1 centred around the Na feature.}
 \label{fig:slc2}
\end{figure*}

\begin{figure*}
 \includegraphics[width=\textwidth]{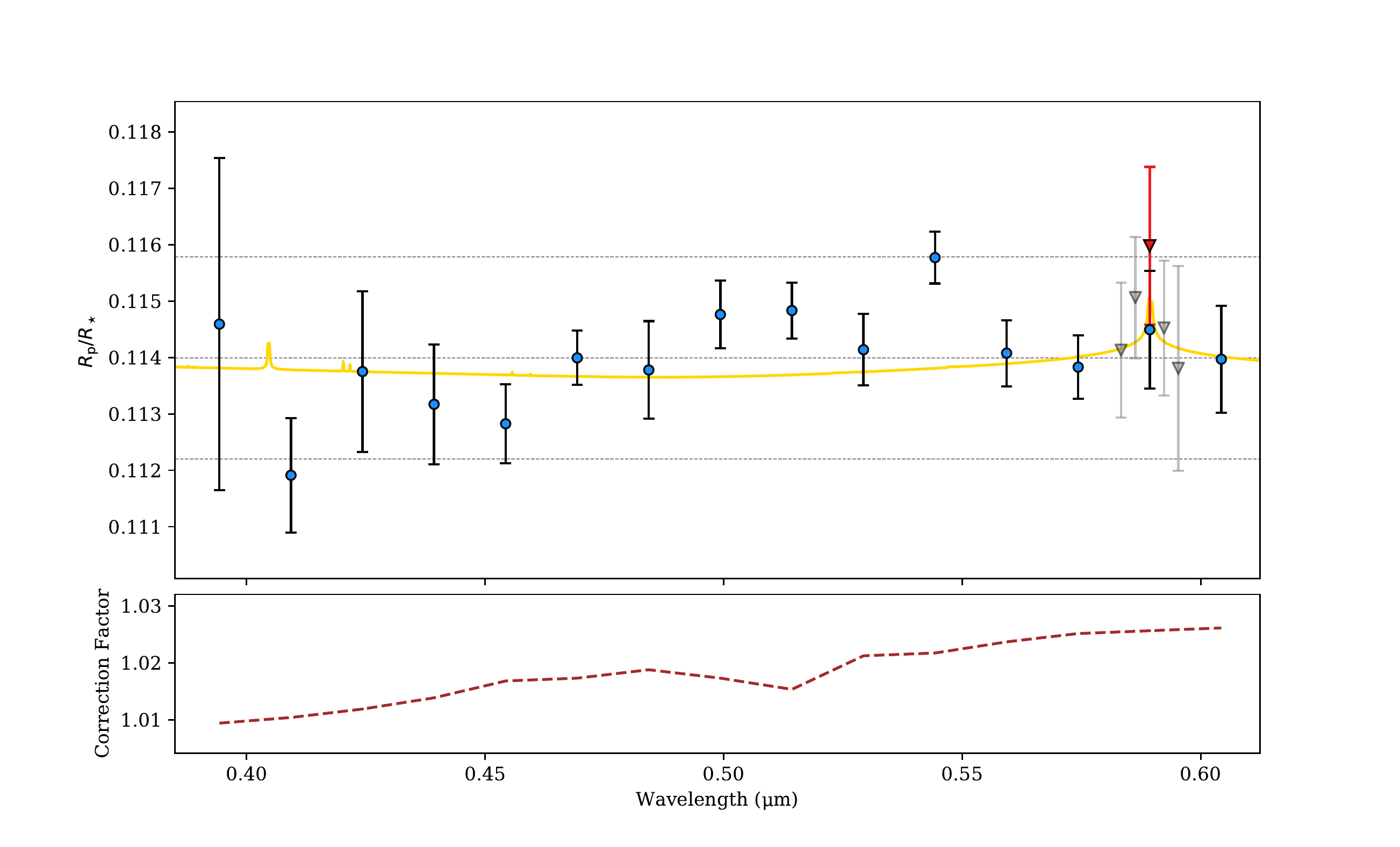}
 \caption{FORS2 transmission spectrum of WASP-103b. The top panel shows the contaminant corrected spectrum. The blue points are the results for the low-resolution light curves. The red triangle is the central high-resolution channel centred on Na and the grey triangles show the high-resolution channels located either side. The grey dashed lines correspond to the mean of the transmission spectrum plus and minus 3 atmospheric scale heights. The gold line shows an example model assuming a clear atmosphere at the terminator. We do not attempt to fit this model to the data, but simply over plot it for reference. The brown dashed line in the lower panel gives an indication of the applied contaminant correction. }
 \label{fig:fors_spec}
\end{figure*}

Enhanced absorption from the alkali metals Na and K has been detected for a range of exoplanets as pronounced features in their transmission spectra \citep[e.g.][]{2002ApJ...568..377C,2011MNRAS.416.1443S,2012MNRAS.422.2477H,2014MNRAS.437...46N,2015A&A...577A..62W,2015MNRAS.446.2428S,2016ApJ...832..191N,2018Natur.557..526N}, whilst many others have shown only partially or highly attenuated features or even completely featureless spectra due to the presence of clouds and/or hazes in the upper atmosphere \citep[e.g.][]{2013MNRAS.436.2974G,2013ApJ...778..183L,2016A&A...587A..67L}. Despite possessing both a high temperature and large radius, WASP-103b nevertheless represents a challenging target for transmission spectroscopy observations, with the amplitude of potential absorption features predicted to be intrinsically small due to its high mass and density. However, signs of enhanced Na and K absorption have previously been observed using Gemini/GMOS by \citet{2017A&A...606A..18L} and we attempted to confirm the Na feature in our FORS2 data by extracting five additional high-resolution light curves in addition to the low-resolution light curves described above.

For our high-resolution channels we used 30\,{\AA} bins with the central bin placed at the mid point of the Na doublet at 5892.9\,{\AA} and two bins either side in the neighbouring continuum. Only Na is covered by the 600B grism and so we are unable to search for additional K absorption using the FORS2 dataset. The resulting light curves show similar systematic features as for the white light curve and low-resolution light curves and so we fitted each of these light curves following the same steps as before, applying the same common-mode correction. The narrow spectral bins used for extraction are shown in Figure 1 and the corresponding light curves in Figure 4.

\subsection{Correction for Contaminant Star}
Previous observations \citep[e.g.][]{2015A&A...579A.129W,2016ApJ...827....8N,2017AJ....153...34C} have revealed that WASP-103 harbours a faint K5V companion star (\textit{T}$_\mathrm{eff}$\,$\approx$\,4400\,$\pm$\,200\,K) within 0.24$''$ and it is likely that this pair are gravitationally bound. Flux contamination from a blended companion has the potential to introduce additional wavelength-dependent effects in both transmission and emission spectra if not properly accounted for \citep[e.g.][]{2012ApJ...760..140C,2016A&A...587A..67L}. At such a small angular separation both of these stars are blended in our observations necessitating a contaminant correction which we implemented as follows: first we obtained theoretical PHOENIX spectra \citep{2013A&A...553A...6H} for WASP-103 and the companion interpolated to the stellar properties reported in \citet{2017AJ....153...34C}. We then estimate the flux contribution integrated over each of our wavelength channels due to the contaminant given by:
\begin{equation}
    \frac{\textit{F}_\mathrm{cont}}{\textit{F}_\mathrm{W103}} = \left(\frac{\textit{R}_\mathrm{cont}}{\textit{R}_\mathrm{W103}}\right)^2\left(\frac{\textit{M}_\mathrm{cont}}{\textit{M}_\mathrm{W103}}\right),
\end{equation}
where \textit{M}$_\mathrm{cont}$ and \textit{M}$_\mathrm{W103}$ are the integrated model fluxes for each passband and \textit{R}$_\mathrm{cont}$/\textit{R}$_\mathrm{W103}$ is the contaminant to target radius ratio. Finally, we used the estimated flux contributions to apply dilution correction factors to each spectral bin in order to account for the contamination and the resulting flux ratios and decontaminated planet-to-star radius ratios are shown in Table 4. Our estimated values are consistent (over the overlapping wavelengths) with those calculated by \citet{2017A&A...606A..18L} who used a similar method to obtain their estimates.
For our analysis we have simply applied the correction factors to our measured radius ratios and uncertainties and have not attempted to account for the small uncertainties which are introduced by these factors. There are a number of other models available which could be used to estimate the flux contamination, the specific choice of which could potentially result in different spectral shapes. \citet{2017A&A...606A..18L} estimated the uncertainties introduced by adopting the PHOENIX model of the K5V companion by carrying out a large number of simulations and found that, while the entire transmission spectrum may be subject to an overall offset of 0.0013 in \textit{R}$_\mathrm{p}$/\textit{R}$_\mathrm{\star}$, any introduced wavelength-dependent slopes are small and therefore unlikely to significantly alter our results even in the most extreme case. The overall effect of our contaminant correction is to add a vertical shift to the transmission spectrum. An indication of the applied contaminant correction is shown in the lower panel of Figure 5.

\begin{table}
 \caption{Calculated flux ratios used to derive the correction factors for each spectral bin and resulting decontaminated transmission spectrum. }
 \label{tab4}
 \begin{tabular}{lcc}
  \hline
  Wavelength  & Flux Ratio & Radius Ratio\\
  Centre [Range] ({\AA}) & \textit{F}$_\mathrm{cont}$/\textit{F}$_\mathrm{W103}$ & \textit{R}$_\mathrm{p}$/\textit{R}$_\mathrm{\star}$  \\
  \hline
  3943 [3868-4018] & 0.019 & 0.11459 $\pm$ 0.00295\\
  4093 [4018-4168] & 0.021 & 0.11191 $\pm$ 0.00102\\
  4243 [4168-4318] & 0.024 & 0.11375 $\pm$ 0.00106 \\
  4393 [4318-4468]& 0.028 & 0.11317 $\pm$ 0.00115\\
  4543 [4468-4618] & 0.034 & 0.11283 $\pm$ 0.00070\\
  4693 [4618-4768] & 0.035 & 0.11399 $\pm$ 0.00048\\
  4843 [4768-4918] & 0.038 & 0.11378 $\pm$ 0.00087\\
  4993 [4918-5068] & 0.035 & 0.11476 $\pm$ 0.00060\\
  5143 [5068-5218] & 0.031 & 0.11483 $\pm$ 0.00050\\
  5293 [5218-5368] & 0.043 & 0.11414 $\pm$ 0.00064\\
  5443 [5368-5518] & 0.044 & 0.11577 $\pm$ 0.00046\\
  5593 [5518-5668] & 0.048 & 0.11408 $\pm$ 0.00059\\
  5743 [5668-5818] & 0.051 & 0.11383 $\pm$ 0.00056\\
  5893 [5818-5968] & 0.052 & 0.11449 $\pm$ 0.00104\\
  6043 [5968-6118] & 0.053& 0.11397 $\pm$ 0.00095\\
  \hline
  central high-resolution channel\\
  5893 [5883-5903] & 0.048 & 0.11598 $\pm$ 0.00140\\
   \hline
 \end{tabular}
\end{table}

\begin{figure*}
 \includegraphics[width=\textwidth]{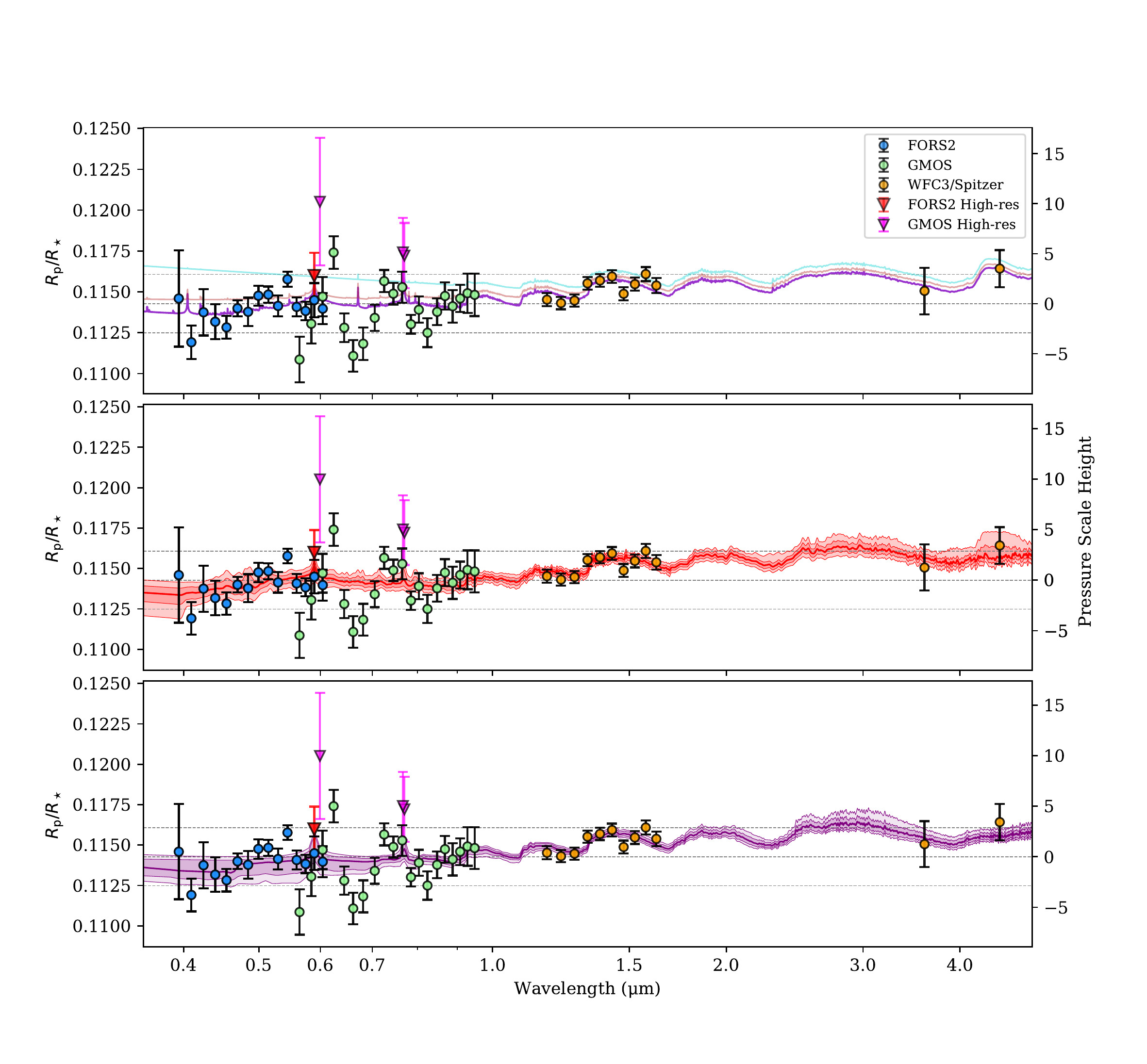}
 \caption{Combined optical-infrared transmission spectrum of WASP-103b obtained from FORS2, Gemini/GMOS, WFC3 and Spitzer observations. The blue points are the results from the low-resolution FORS2 light curves. The green and orange points are the results from \citet{2017A&A...606A..18L} and \citet{2018AJ....156...17K} respectively. The red triangles show the result for our high-resolution bin centred on the Na feature. The magenta triangles show the high-resolution Na and K measurements from \citet{2017A&A...606A..18L} (Na measurement slightly offset for clarity). The top panel shows the best-fit model (purple line) from our forward modelling along with the best-fit full cloud (crimson) and full haze (turquoise) models for comparison. Both of these models have been slightly offset from the best-fit for clarity. The middle panel shows the median fit (red line) from the retrieval analysis using AURA along with the 1 and 2 sigma significance contours (red/light red). The lower panel shows the median fit (purple line) from the retrieval analysis using NEMESIS along with the 1 and 2 sigma significance contours (purple/light purple). For all panels the dashed lines correspond to the mean of the transmission spectrum plus and minus 3 atmospheric scale heights.}
\end{figure*}

Our decontaminated FORS2 transmission spectrum for WASP-103b is shown in the top panel of Figure 5. Our results are consistent with a linear fit to the transmission spectrum with a $\chi$$^2$ value of 19.62 for 13 degrees of freedom or reduced $\chi$$^2$ of 1.51, and we calculate a low significance for the Na feature ($<$\,1.5\,$\sigma$). We discuss these results further in Section 5.

\section{Atmospheric Modelling}

\subsection{Goyal forward models}

We combined the optical data from the FORS2 and GMOS observations with those obtained by \citet{2018AJ....156...17K} using WFC3 and Spitzer in the near-IR to produce a complete, optical-infrared transmission spectrum of WASP-103b. Our system scale and inclination parameters are fixed to those assumed in the WFC3/Spitzer analysis and we apply a small offset to the GMOS spectrum ($\approx$\,--\,4\,$\times$\,10$^{-6}$\,\textit{R}$_\mathrm{p}$/\textit{R}$_\mathrm{\star}$) calculated using the overlapping FORS2/GMOS region to stitch the spectra together in the vertical direction, accounting for the difference in white light curve parameters. We compared the full transmission spectrum with a generic grid of forward models generated using the one-dimensional radiative-convective equilibrium code ATMO \citep[][]{Amundsen_2014,dr02400q,go01600j,2015ApJ...804L..17T,2016ApJ...817L..19T}. Each model in the grid assumes chemical equilibrium abundances and isothermal pressure-temperature (p-T) profiles with the entire grid exploring 24 equilibrium temperatures from 300\,--\,2600\,K in steps of 100\,K, six metallicities (0.1\,--\,200\,x\,solar), four planetary gravities (5\,--\,50\,ms$^{-2}$), four C/O ratios (0.35\,--\,1.0) and four parameters each describing scattering hazes and uniform clouds. The haze parameter defines the wavelength dependent Rayleigh scattering profile for small particles whilst the cloud parameter defines the uniform grey scattering profile, simulating the effects of a cloud deck from 0 to 100$\%$ cloud opacity across all wavelengths.
Each of the models considers H$_2$\,-\,H$_2$, H$_2$\,-\,He collision-induced absorption (CIA) and opacities due to H$_2$O, CO$_2$, CO, CH$_4$, NH$_3$, Na, K, Li, Rb, Cs, TiO, VO, FeH, PH$_3$, H$_2$S, HCN, SO$_2$ and C$_2$H$_2$. The source of these opacities and their pressure broadening parameters can be found in \citet{Amundsen_2014} and  \citet{go01600j} and we adopt the Na and K pressure broadened line profiles from \citet{Burrows_2000}. 

The generic model grid is baselined for a Jupiter radius planet around a Solar radius star and each model in the grid can then be scaled based on the planetary radius, stellar radius and surface gravity of WASP-103b using the scaling relationship derived in \citet{go01600j}. We note that since we fit for the temperature, we don't scale to the planetary equilibrium temperature in the equation (i.e. the temperature terms cancel out). The generic model grid has been developed for two different condensation schemes: `local' and `rainout' condensation. In the `local condensation' scheme each model level is independent, with the chemical composition depending only on elemental abundances and local conditions of pressure and temperature. In this scheme, any condensates which form will deplete elements only within that layer of the atmosphere. In the `rainout' scenario, condensates which form will also deplete elements in all layers above, in addition to the local layer, and the chemical abundances of each layer will therefore also be dependent on all other deeper layers of the atmosphere.

Our best fit model is found after scaling to the parameters of WASP-103b and using a least square minimization procedure with free vertical offset in \textit{R}$_\mathrm{p}$/\textit{R}$_\mathrm{\star}$ (see \citealt{go01600j} for further details of the grid parameters and implementation).
Here we only consider the rainout condensation scenario, although we note that using the local condensation approach made little difference to our interpretation of the observed spectrum. Our best-fit model is shown in the top panel of Figure 6 along with full cloud/full haze models for comparison. Our best-fit model favours a clear atmosphere with \textit{T}\,=\,1700\,K, super-solar metallicity [M/H]\,=\,+\,1.7, solar C/O ratio [C/O]\,=\,0.56 and planetary gravity \textit{g}\,=\,10\,m s$^{-2}$ corresponding to a $\chi^2$ of 81.46 for 44 degrees of freedom or reduced $\chi^2$ of 1.89. Our model shows evidence of H$_2$O at 1.4 microns and no evidence of either clouds or hazes. We obtain a reduced $\chi^2$ of 3.01 and 1.99 for the full haze and full cloud models respectively (note that the models shown in the top panel of Figure 6 have been subsequently, and arbitrarily, shifted for clarity).

\subsection{Atmospheric Retrieval with AURA}

In addition to the forward modelling described above we performed a Bayesian atmospheric retrieval on the full dataset to try to constrain the atmospheric composition and temperature structure at the day-night terminator of WASP-103b. Our retrieval uses an adaptation of the retrieval code AURA \citep[]{Pinhas_2018,2019AJ....157..206W}, with the method having already been successfully implemented for a number of transmission spectra \citep[e.g.][]{Pinhas_2018,2019MNRAS.482.1485P,2019AJ....157..206W,Madhusudhan_2020}. Our atmospheric retrieval code consists of two components: a forward model to predict the atmospheric spectrum and an algorithm for statistical parameter estimation. The model solves line-by-line radiative transfer for a transmission geometry and assumes hydrostatic equilibrium, a plane parallel atmosphere and uniform volume mixing ratios. In addition to CIA we include the sources of chemical opacity expected to be prominent in hot-Jupiter atmospheres in the observed spectral range: H$_2$O, Na, K, TiO, VO, AlO, HCN, CO and CO$_2$. Cross sections for these sources are calculated by \citet{Gandhi_2017} from a range of databases \citep[]{2010JQSRT.111.2139R,2013JQSRT.130....4R,2012JQSRT.113.1276R,2016JMoSp.327...73T} including EXOMOL \citep[]{2016JMoSp.327...73T,Yurchenko_2011,Barber_2013,Yurchenko_2014}, HITEMP \citep{2010JQSRT.111.2139R} and HITRAN \citep{2012JQSRT.113.1276R}. The model uses a parameterised, one-dimensional p-T profile with the atmosphere divided into three distinct zones defined by pressure values $P_{\rm 1}$, $P_{\rm 2}$ and $P_{\rm 3}$. $T_{\rm 0}$ (in Kelvin) defines the temperature at the top of the atmosphere, while $\alpha_{\rm 1}$ and $\alpha_{\rm 2}$ describe the gradient of the profile. The model also includes the a priori unknown reference pressure $P_{\rm ref}$ at $R_{\rm p}$. Our model also considers the contributions from homogeneous/inhomogeneous cloud/haze coverage \citep{2017MNRAS.469.1979M}. This is parameterised with a cloud-deck altitude ($P_{\rm cloud}$, in bars), a Rayleigh enhancement factor ($a$, a linear scaling of the opacity from H$_2$ Rayleigh scattering), and $\gamma$, which describes the Rayleigh scatting slope. Finally, we include a term $\bar\phi$, which describes the terminator averaged cloud/haze contribution. This varies between 0 (clear atmosphere) and 1 (fully cloudy atmosphere), where the forward model is computed both with and without the cloud/haze model before being averaged (with weighting governed by $\bar\phi$). For more details on these parameters and a detailed description of the method see \citet{Pinhas_2018}.

A statistical sampling algorithm is used to infer the properties of the exoplanetary atmosphere - i.e. the posterior distributions of the forward model parameters and their credibility intervals. For our retrieval we used the Nested Sampling algorithm MultiNest \citep{Feroz_2009} implemented with PyMultiNest \citep{Buchner_2014} which, in addition to robust parameter estimation, also allows calculation of the Bayesian evidence term $\mathcal{Z}$, facilitating model comparison and the calculation of detection significances. Our abundances are presented as the average terminator volume mixing ratios $X_{\rm i}$\,=\,$n_{\rm i}$/$n_{\rm tot}$. The normalised abundance is relative to a `solar' value - that expected in equilibrium at the relevant temperature for an atmosphere with solar elemental abundances \citep[e.g.][]{Asplund_2009,2012ApJ...758...36M}. We employed a uniform prior between 800\,--\,2800\,K for the temperature at the top of the atmosphere. We follow a similar procedure to that outlined in \citet{Pinhas_2018} and set an upper limit on the prior for $T_{\rm 0}$ to be a few 100\,K above the equilibrium temperature (\textit{T}$_\mathrm{eq}$\,$\approx$\,2500\,K). The reason for this restriction is that the value for $T_{\rm 0}$ is expected to be significantly below the equilibrium temperature, and allowing for much higher values can lead to unphysical solutions. (see Table A1 in the appendix for the full list of our assumed priors). We also include an offset between the datasets as a free parameter in our retrieval to account for any remaining discrepancies in the overall levels.

Our retrieval indicates a detection of H$_2$O with a significance of 4.0\,$\sigma$. We retrieve a terminator H$_2$O abundance of log(\textit{X}$_\mathrm{H_2O}$) = --1.73$_\mathrm{-0.55}^\mathrm{+0.38}$, corresponding to a $\sim$\,40\,$\times$ solar abundance composition. We also constrain the Na abundance to log(\textit{X}$_\mathrm{Na}$) = --3.02$^\mathrm{+0.98}_\mathrm{-2.43}$ though with a low significance (2.0\,$\sigma$) and constrain the terminator-averaged cloud/haze fraction $\Bar{\phi}$ to be 0.35$_\mathrm{-0.15}^\mathrm{+0.15}$. The retrieved H$_2$O and Na abundance estimates, solar-relative abundances and detection significances are listed in Table 5 and the median fit along with the 1 and 2 sigma confidence contours are shown in the middle panel of Figure 6. The complete atmospheric retrieval results including marginalised posterior probability densities are presented in the appendix.

\begin{table}
 \caption{Retrieved terminator H$_2$O and Na abundances, solar-normalised abundances and detection significances using AURA. }
 \label{tab5}
 \begin{tabular}{lccc}
  \hline
  Species & Abundance  & Normalised Abundance & Significance\\
  \hline
  H$_2$O & -1.73$_\mathrm{-0.55}^\mathrm{+0.38}$ & 40 & 4.0-$\sigma$\\
  Na & -3.02$_\mathrm{-2.43}^\mathrm{+0.98}$ & 600 & 2.0-$\sigma$\\
     \hline
 \end{tabular}
\end{table}

\subsection{Atmospheric Retrieval with NEMESIS}

We also compared the results from our AURA retrieval to those obtained using the NEMESIS radiative transfer and retrieval algorithm. NEMESIS (Non-linear optimal Estimator for MultivariatE spectral analySIS) was originally designed to model Solar System objects \citep{IRWIN20081136} but has since been modified for exoplanet atmospheres \citep[e.g.][]{Lee_2011,Barstow_2013,Lee_2014,Barstow_2016,Bruno_2019,Barstow_2020}. In its original form NEMESIS paired a fast correlated-k \citep{doi:10.1029/90JD01945} forward model with an efficient optimal estimation algorithm for parameter estimation \citep{doi:10.1142/3171}, though more recently this has been upgraded to take advantage of the PyMultiNest algorithm \citep{Krissansen_Totton_2018}, allowing for full exploration of non-Gaussian posterior distributions. As with our AURA retrieval we also include CIA and the sources for our opacities are found in various databases including those from \citet{osti_5642348,osti_6904948}, \citet{borysowfm89,borysow97}, \citet{borysow02}, EXOMOL \citep{Chubb2020}, and NIST \citep{NaKlinelist}. We assume an isothermal p-T profile and represent clouds with the cloud top and base pressures and index of the scattering slope as free parameters in the model, along with the total optical depth. We use similar prior ranges as for the AURA retrieval.

We find our results from NEMESIS are in excellent agreement with those obtained from AURA, with the log opacity of -3.98$_\mathrm{-3.70}^\mathrm{+3.72}$ being consistent with a clear atmosphere and terminator abundances for H$_2$O and Na of log(\textit{X}$_\mathrm{H_2O}$) = --1.33$_\mathrm{-0.58}^\mathrm{+0.22}$ and log(\textit{X}$_\mathrm{Na}$) = --3.25$_\mathrm{-4.34}^\mathrm{+1.43}$ respectively. The retrieved abundance estimates agree within 1\,$\sigma$ with the corresponding AURA estimates, giving us further confidence in the model fits. Our best-fit retrieval model along with 1 and 2 sigma significance contours are shown in the lower panel of Figure 6 and the complete retrieval results including marginalised posterior probability densities are given in the appendix.

\section{Discussion}

WASP-103b is an ultra-short period hot-Jupiter with a mass and radius of 1.49 \textit{M}$_\mathrm{J}$ and 1.53 \textit{R}$_\mathrm{J}$ respectively, it has an equilibrium temperature close to 2500\,K, a surface gravity $\approx$\,15\,m\,s$^{-2}$ and an atmospheric scale height $\approx$\,600\,km or 0.0006 \textit{R}$_\mathrm{p}$/\textit{R}$_\mathrm{\star}$. \citet{2015MNRAS.447..711S} observed a strong (7.3\,$\sigma$) wavelength-dependent slope in the optical and found that this held even after applying a correction for the contaminant star \citep{2016MNRAS.463...37S}. Conversely, \citet{2017A&A...606A..18L} observed signs of strong absorption in the cores of both the alkali features using Gemini/GMOS but did not recover the V-shaped pattern or any evidence for a Rayleigh scattering signature. A study by \citet{2018AJ....156...17K} revealed a featureless nightside-corrected transmission spectrum which was consistent with a flat line fit within 1\,$\sigma$ across the WFC3 and Spitzer bands. Additionally, they determined that their phase-resolved spectra were consistent with blackbody emission and attributed the lack of detection of H$_2$O  features to partial dissociation.

Several recent studies \citep[e.g.][]{2019A&A...626A.133H,2019A&A...631A..79H} suggest that the large temperature gradients expected for ultra-hot Jupiters such as WASP-103b likely lead to cloud-free daysides but rather cloudy nightsides. Furthermore, the regions probed by transmission spectroscopy observations may not be homogeneously cloudy and can also feature strong morning/evening terminator asymmetries, with similar asymmetries expected for the amount of observable gas in these regions. 

\citet{Staab_2016} estimated the log(\textit{R$'$}$_\mathrm{HK}$) value for WASP-103 finding it to be $\approx$\,4.57 which was higher than expected from the system age. However, despite this activity we don't expect the corresponding stellar heterogeneity to result in a measurable offset in our transmission spectrum given the spectral type (F8V) of the host star. In \citet{2019AJ....157...96R} the estimated contamination for an F8 dwarf is a factor of $\sim$\,1.001 to $\sim$\,1.002, corresponding to an offset in transit depth of $\sim$\,0.0025\% which is within the error of our measurements.

\subsection{FORS2 Transmission Spectrum}

Our decontaminated FORS2 transmission spectrum is shown in Figure 5 along with an example model which assumes a clear atmosphere at the terminator. The horizontal lines show the weighted average and plus and minus three scale heights, with one scale height corresponding to $\approx$\,600\,km or 1\,$\times$\,10$^{-3}$ in transit depth. Our uncertainties range from $\approx$\,5\,$\times$\,10$^{-4}$ at the centre of the grism to $\ga$\,2\,$\times$\,10$^{-3}$ at the edges showing that our applied common-mode correction is most accurate over the central wavelength bands. We performed a least squares fit for a horizontal line using a Levenburg-Marquardt algorithm and calculate a $\chi$$^2$ of 25.18 for 14 degrees of freedom or reduced $\chi$$^2$ of 1.80 for the decontaminated transmission spectrum. We find an improved fit for a linear model including an upwards slope with a $\chi$$^2$ of 19.62 for 13 degrees of freedom or reduced $\chi$$^2$ of 1.51 and so we are unable to reject a featureless model. In both cases the major contribution to the $\chi$$^2$ value stems from a single outlier at $\sim$\,5500\,{\AA} which does not correspond to absorption from any of the species considered in our retrieval. Masking this single point in our calculation results in a reduced $\chi$$^2$ close to unity. We calculate the significance of the central high-resolution Na measurement to be $<$\,1.5\,$\sigma$. To facilitate a more direct comparison to the results from \citet{2017A&A...606A..18L} we also extracted an additional high-resolution channel centred on the Na feature with a bin width of 20\,{\AA} (the smallest bin width used in their study). Whilst we do observe a slightly higher value for the planet-to-star radius ratio when compared to our 30\,{\AA} bin, we also recover a correspondingly larger uncertainty and calculate a significance of $\approx$\,1.2\,$\sigma$. We therefore conclude that we do not detect strong evidence of Na absorption in our FORS2 dataset. Additionally, we find no evidence for a wavelength-dependent slope towards the blue which would also appear to rule out the strong Rayleigh scattering signature previously inferred. Our measured value for the high-resolution Na channel is lower than that reported by \citet{2017A&A...606A..18L}, though the measurements agree within their 1\,$\sigma$ uncertainties. The significance of our measurement is also slightly reduced (down from $\approx$\,1.7\,$\sigma$ for GMOS to $\approx$\,1.2\,$\sigma$ for FORS2). One possible explanation for this systematic offset in the narrow-band feature is the different methods used to treat the instrumental systematics. In their analysis, \citet{2017A&A...606A..18L} used parametric baseline models along with a common noise model to account for the systematic effects and they performed model selection via the Bayesian information criterion (BIC) to optimally choose from amongst the possible models.

The absence of a pressure broadened Na feature in our FORS2 transmission spectrum could most easily be explained by the presence of a high-altitude cloud deck which acts to obscure the lower regions of the planetary atmosphere and reduces the strength of the spectral features. However, neither the results of our forward modelling nor our retrieval analyses strongly support this conclusion, with each favouring a relatively clear atmosphere at the terminator. Our retrieved value ($\Bar{\phi} = 0.35_\mathrm{-0.15}^\mathrm{+0.15}$) for the cloud/haze fraction parameter using AURA is not consistent with a thick cloud deck acting as a grey absorber, though it perhaps indicates some degree of patchy or inhomogeneous clouds/hazes which may still contribute to the muting of spectral features in the FORS2 spectrum. Furthermore, due to a high mass and density, the amplitude of potential absorption features in the atmosphere of WASP-103 is predicted to be intrinsically small. Therefore the most likely explanation is that a combination of inadequate signal-to-noise and a relatively small atmospheric scale height makes the robust detection of Na in the atmosphere of WASP-103b challenging to obtain from the ground. A further possible explanation is that with such high temperatures Na is largely ionised in the part of the atmosphere that dominates the transmission light curves, leading to low abundances and a small Na feature. This scenario is similar to that found for WASP-18b \citep{2019A&A...626A.133H} where the low-pressure part of the terminators show more Na$^+$ then Na. In principle this could also be the case for K. A final possibility is that Na is not present in the atmosphere, though given its equilibrium temperature ($\approx$\,2500\,K) this explanation is less likely. Higher signal-to-noise observations would be useful to definitively choose between these possible scenarios whilst higher resolution observations will be required to detect the Na core if clouds are present in the atmosphere. With sufficient signal-to-noise (e.g. with MIRI on JWST) it may be possible to verify the presence of cloud species in the mid-infrared where the cloud spectral signatures are more distinct, and perhaps even put constraints on their composition \citep[e.g.][]{2018AJ....155...29W}.

\subsection{Combined transmission spectrum}

 Figure 6 shows our full optical-infrared transmission spectrum for WASP-103b incorporating the FORS2, GMOS, WFC3 and Spitzer data. Our retrieval analyses indicate a detection of H$_2$O absorption in the near-IR with a significance of 4.0\,$\sigma$. All of our best-fit models favour a relatively clear atmosphere at the terminator region. Our best-fit forward model has greater than solar metallicity ([M/H] = +1.7) and solar C/O ratio ([C/O] = 0.56). The retrieved temperature using AURA ($\sim$\,2300\,K) is significantly higher than the best fit temperature from the Goyal grid ($\sim$\,1700\,K). One explanation for this is that for higher temperatures the Goyal grid predicts prominent TiO/VO absorption features in the optical and these features are not well matched to the observed FORS2 and GMOS datasets. Therefore the model tends to favour the highest temperature in the grid which does not result in strong TiO/VO absorption. Another possible explanation is that the Goyal grid uses isothermal p-T profiles whilst the AURA retrieval uses a parameterised profile. The majority of carbon bearing species in the atmosphere are covered by the Spitzer observations and the significant uncertainties in these measurements means that we are unable to put any meaningful constraints on the carbon abundance, with the low retrieved abundances for CO/CO$_2$ possibly reflecting the lack of data in the relevant parts of the spectrum. It is important to acknowledge that the differing paradigms used in the modelling, i.e. equilibrium models vs free chemistry, lead to differing assumptions about the physical and chemical properties of the atmosphere. On the one hand equilibrium models may lack the full flexibility required to model the wide range of exoplanetary atmospheres which could deviate significantly from equilibrium assumptions \citep{Madhusudhan_2018}, whilst on the other it is possible that the free chemistry approach could result in some un-physical combinations of parameters.
 
 In their analysis \citet{2018AJ....156...17K} found that their nightside-corrected transmission spectrum was consistent with a flat line fit at the 1\,$\sigma$ level across the WFC3 and Spitzer bands and concluded that they did not detect any evidence of H$_2$O absorption. However, in that analysis they also found that their corrected spectrum was consistent with predictions from a general circulation model including H$_2$O features in the WFC3 bandpass and suggested that further high-precision observations might render these features detectable. One potential explanation for our contrasting result is that \citet{2018AJ....156...17K} only had access to the WFC3 and Spitzer data and their analysis did not include the additional optical data from FORS2 and GMOS. Having access to the full optical-infrared data is important for obtaining accurate retrievals as precise modelling of the continuum is crucial for breaking degeneracies between model parameters \citep[]{Pinhas_2018,2018AJ....155...29W}. Furthermore, our retrieval method also includes the possibility of patchy cloud coverage which can affect the shape of the H$_2$O feature in the WFC3 bandpass depending on the degree of cloud coverage \citep[e.g.][]{2016ApJ...820...78L}. We calculated the $\chi$$^2$ value for a featureless (flat) fit to the WFC3 data alone finding a reduced $\chi$$^2$ >\,2. We therefore conclude that our retrieval results provide reasonable evidence for H$_2$O absorption at the day-night terminator of WASP-103b. \citet{2018AJ....156...17K} also demonstrated that WASP-103b shows poor heat redistribution between the day and night sides and so one possible explanation as to why they did not detect H$_2$O in their phase-curve observations is that the majority of H$_2$O on the dayside is thermally dissociated but still exists with measurable abundance on the cooler nightside and terminator of the planet. For example, this is the case for the similar ultra-hot Jupiter WASP-121b, which shows H$_2$O absorption at the day-night terminator \citep[]{Evans_2016,2018AJ....156..283E} and the same feature in emission on the dayside hemisphere \citep[]{Evans_2017}, although significantly weakened due to thermal dissociation \citep[]{Parmentier_2018,Mikal_Evans_2019}.

 We do not obtain particularly strong evidence for Na in our retrievals. However, our detection significance of 2.0\,$\sigma$, along with the previous Na detection in the GMOS band, means that neither can we definitively rule out its presence.  
 
 Finally, we should also note that a number of studies have recently highlighted some of the challenges inherent for one-dimensional retrievals of ultra-hot Jupiters: in \citet{Pluriel_2020} they show that thermal dissociation and the strong day to night temperature gradient lead to a chemical composition dichotomy between the two hemispheres which can strongly bias retrieved abundances. \citet{2020ApJ...893L..43M} demonstrates that one-dimensional retrievals can significantly underestimate the temperature - particularly for ultra-hot Jupiters - and resulted in an overestimate of the H$_2$O abundance and an underestimate in the H$^-$ abundance, whilst in \citet{Irwin_2020} they found that their 2.5-dimensional retrieval approach was more reliable for modelling phase curve observations of exoplanets compared to the one-dimensional approach.

\section{Conclusion}

We have presented ground-based FORS2 observations of the highly irradiated hot-Jupiter WASP-103b covering one full transit and extracting a transmission spectrum over the range $\approx$\,400\,--\,600\,nm using the technique of differential spectrophotometry. We used a Gaussian process to simultaneously model the deterministic transit component and the instrumental systematics avoiding the need to specify a specific functional form for the systematics. We used the derived systematics model to correct our spectroscopic light curves using a common-mode correction to improve the precision of our transmission spectrum, reaching a typical precision of $\approx$\,2\,$\times$\,10$^{-4}$ in transit depth. We accounted for flux contamination due to a blended companion by applying a dilution correction factor to each spectral bin based on an estimate of the flux ratios derived from PHOENIX model spectra of WASP-103 and the contaminant star. 

Our analysis of the FORS2 data reveals a featureless spectrum across the full range of the observations and we find no evidence for either alkali metal absorption or Rayleigh scattering. The featureless FORS2 transmission spectrum and absence of broad absorption features could most easily be explained by either a low abundance of Na in the atmosphere, or the presence of optically thick, high-altitude clouds or other aerosols which cause broad-band extinction, masking the absorption signatures in the upper atmosphere either by scattering or absorption across the full range of the observations.

To investigate this further we fit a grid of forward models and performed an atmospheric retrieval on the full optical-infrared spectrum incorporating the additional observations from GMOS, WFC3 and Spitzer. Our retrieval indicates a detection of H$_2$O at the 4.0\,$\sigma$ level and Na at the lower significance of 2.0\,$\sigma$. We compared the results from our AURA retrieval with those obtained using NEMESIS finding excellent agreement between the two approaches. In all cases we find that a relatively clear atmosphere at the terminator provides the best fit to our data. We conclude that the most likely explanation for our featureless FORS2 spectrum is due to a combination of low signal-to-noise and the inherently small scale height of WASP-103b, although patchy/inhomogeneous clouds or hazes may still play a part in damping the absorption features in the optical. Additional observations at high signal-to-noise might be able to resolve the Na feature whilst high-resolution observations will be required to detect the narrow Na core if clouds/hazes are present in the atmosphere of WASP-103b. 

\section*{Acknowledgements}
We are extremely grateful to the anonymous referee for careful reading of the manuscript. This work is based on observations collected at the European Organisation for Astronomical Research in the Southern Hemisphere under ESO programme 199.C-0467. J.W. would like to acknowledge funding from the Northern Ireland Department for the Economy. N.P.G. gratefully acknowledges support from Science Foundation Ireland and the Royal Society in the form of a University Research Fellowship. A.L.C. is funded by an STFC studentship. We thank Patrick Irwin for the use of NEMESIS. We are grateful to the developers of the NumPy, SciPy, Matplotlib, iPython and Astropy packages, which were used extensively in this work \citep{jones2001scipy,2007CSE.....9...90H,2007CSE.....9c..21P,2013A&A...558A..33A,2020SciPy-NMeth}.

\section*{Data availability}
The data in this article are available from the ESO Science Archive Facility (\url{http://archive.eso.org}) with program ID 199.C-0467. The data products generated from the raw data are available upon request from the author.

\bibliographystyle{mnras}
\bibliography{ref}

\appendix
\section{Complete Atmospheric Retrieval Results using AURA and NEMESIS}

\begin{table}
    \caption{Retrieved atmospheric parameters using AURA.}
    \label{paramstable}
    \begin{tabular}{l|c|c}
        Parameter & Prior & Value \\
        \hline
        $\mathrm{log}(X_{\mathrm{H_2O}})$ & $\mathcal{U}(-12, -1)$ & $-1.73 ^{+ 0.38 }_{- 0.55 }$\\
        $\mathrm{log}(X_{\mathrm{Na}})$ & $\mathcal{U}(-12, -1)$ & $-3.02 ^{+ 0.98 }_{- 2.43 }$\\
        $\mathrm{log}(X_{\mathrm{K}})$ & $\mathcal{U}(-12, -1)$ & $-6.70 ^{+ 2.66 }_{- 3.06 }$\\
        $\mathrm{log}(X_{\mathrm{TiO}})$ & $\mathcal{U}(-12, -1)$ & $-9.59 ^{+ 1.38 }_{- 1.41 }$\\
        $\mathrm{log}(X_{\mathrm{VO}})$ & $\mathcal{U}(-12, -1)$ & $-9.25 ^{+ 1.38 }_{- 1.58 }$\\
        $\mathrm{log}(X_{\mathrm{AlO}})$ & $\mathcal{U}(-12, -1)$ & $-9.47 ^{+ 1.42 }_{- 1.52 }$\\
        $\mathrm{log}(X_{\mathrm{HCN}})$ & $\mathcal{U}(-12, -1)$ & $-6.57 ^{+ 3.33 }_{- 3.27 }$\\
        $\mathrm{log}(X_{\mathrm{CO}})$ & $\mathcal{U}(-12, -1)$ & $-6.81 ^{+ 3.05 }_{- 3.15 }$\\
        $\mathrm{log}(X_{\mathrm{CO_2}})$ & $\mathcal{U}(-12, -1)$ & $-6.72 ^{+ 3.04 }_{- 3.16 }$\\
        $T_{0}$ & $\mathcal{U}(800, 2800)$  & $2293 ^{+ 283 }_{- 266 }$\\
        $\alpha_{1}$ & $\mathcal{U}(0.02, 2.0)$ & $1.17 ^{+ 0.50 }_{- 0.50 }$\\
        $\alpha_{2}$ & $\mathcal{U}(0.02, 2.0)$ & $1.12 ^{+ 0.52 }_{- 0.53 }$\\
        $\mathrm{log}(P_1)$[bar] & $\mathcal{U}(-6, 2)$ & $-1.75 ^{+ 1.56 }_{- 1.56 }$\\
        $\mathrm{log}(P_2)$[bar] & $\mathcal{U}(-6, 2)$ & $-4.08 ^{+ 1.60 }_{- 1.23 }$\\
        $\mathrm{log}(P_3)$[bar] & $\mathcal{U}(-2, 2)$ & $0.53 ^{+ 0.94 }_{- 1.25 }$\\
        $\mathrm{log}(P_{\mathrm{ref}})$[bar] & $\mathcal{U}(-6, 2)$  & $0.99 ^{+ 0.61 }_{- 0.74 }$\\
        $\mathrm{log}(a)$ & $\mathcal{U}(-4, 10)$ & $5.01 ^{+ 2.73 }_{- 5.41 }$\\
        $\gamma$ & $\mathcal{U}(-20, 2)$  & $-9.70 ^{+ 7.16 }_{- 6.37 }$\\
        $\mathrm{log}(P_\mathrm{cloud})$[bar] & $\mathcal{U}(-6, 2)$ & $-3.08 ^{+ 2.85 }_{- 1.61 }$\\
        $\bar{\phi}$ & $\mathcal{U}(0, 1)$  & $0.35 ^{+ 0.15 }_{- 0.15 }$\\
        \hline
    \end{tabular}
\end{table}

\begin{table}
    \caption{Retrieved atmospheric parameters using NEMESIS.}
    \label{paramstable}
    \begin{tabular}{l|c|c}
        Parameter & Prior & Value \\
        \hline
        $\mathrm{log}(X_{\mathrm{H_2O}})$ & $\mathcal{U}(-11, -1)$ & $-1.33 ^{+ 0.22 }_{- 0.58 }$\\
        $\mathrm{log}(X_{\mathrm{Na}})$ & $\mathcal{U}(-11, -1)$ & $-3.25 ^{+ 1.43 }_{- 4.34 }$\\
        $\mathrm{log}(X_{\mathrm{K}})$ & $\mathcal{U}(-11, -1)$ & $-4.77 ^{+ 1.62 }_{- 2.73 }$\\
        $\mathrm{log}(X_{\mathrm{TiO}})$ & $\mathcal{U}(-13, -1)$ & $-9.46 ^{+ 1.64 }_{- 2.21 }$\\
        $\mathrm{log}(X_{\mathrm{VO}})$ & $\mathcal{U}(-13, -1)$ & $-10.63 ^{+ 1.55 }_{- 1.44 }$\\
        $\mathrm{log}(X_{\mathrm{AlO}})$ & $\mathcal{U}(-13, -1)$ & $-10.00 ^{+ 2.17 }_{- 1.89 }$\\
        $\mathrm{log}(X_{\mathrm{H-}})$ & $\mathcal{U}(-13, -1)$ & $-6.97 ^{+ 3.65 }_{- 3.78 }$\\
        $\mathrm{log}(X_{\mathrm{e-}})$ & $\mathcal{U}(-13, -1)$ & $-6.95 ^{+ 3.61 }_{- 3.82 }$\\
        $\mathrm{log}(X_{\mathrm{H}})$ & $\mathcal{U}(-13, -1)$ & $-7.26 ^{+ 3.76 }_{- 3.61 }$\\
        \textit{R}$_\mathrm{J}$ & $\mathcal{U}(1, 2.2)$  & $1.57 ^{+ 0.01 }_{- 0.02 }$\\
        \textit{T}$_\mathrm{strat}$ &   $\mathcal{U}(500, 3000)$& $1721.65 ^{+ 443.06 }_{- 264.84 }$\\
        $\mathrm{log}(P_\mathrm{top})$[atm] & $\mathcal{U}(-8, 1)$ & $-4.57 ^{+ 2.47 }_{- 2.08 }$\\
        $\mathrm{log}(P_\mathrm{base})$[atm] & $\mathcal{U}(\mathrm{Cloud\,top}, 1)$ & $-0.68 ^{+ 1.10 }_{- 1.83 }$\\
        Scattering index & $\mathcal{U}(0, 14)$  & $7.62 ^{+ 4.07 }_{- 4.48 }$\\
        {log}(Opacity) & $\mathcal{U}(-10, 20)$ & $-3.98 ^{+ 3.72 }_{- 3.70 }$\\
        \hline
    \end{tabular}
\end{table}

\begin{figure*}
 \includegraphics[width=\textwidth]{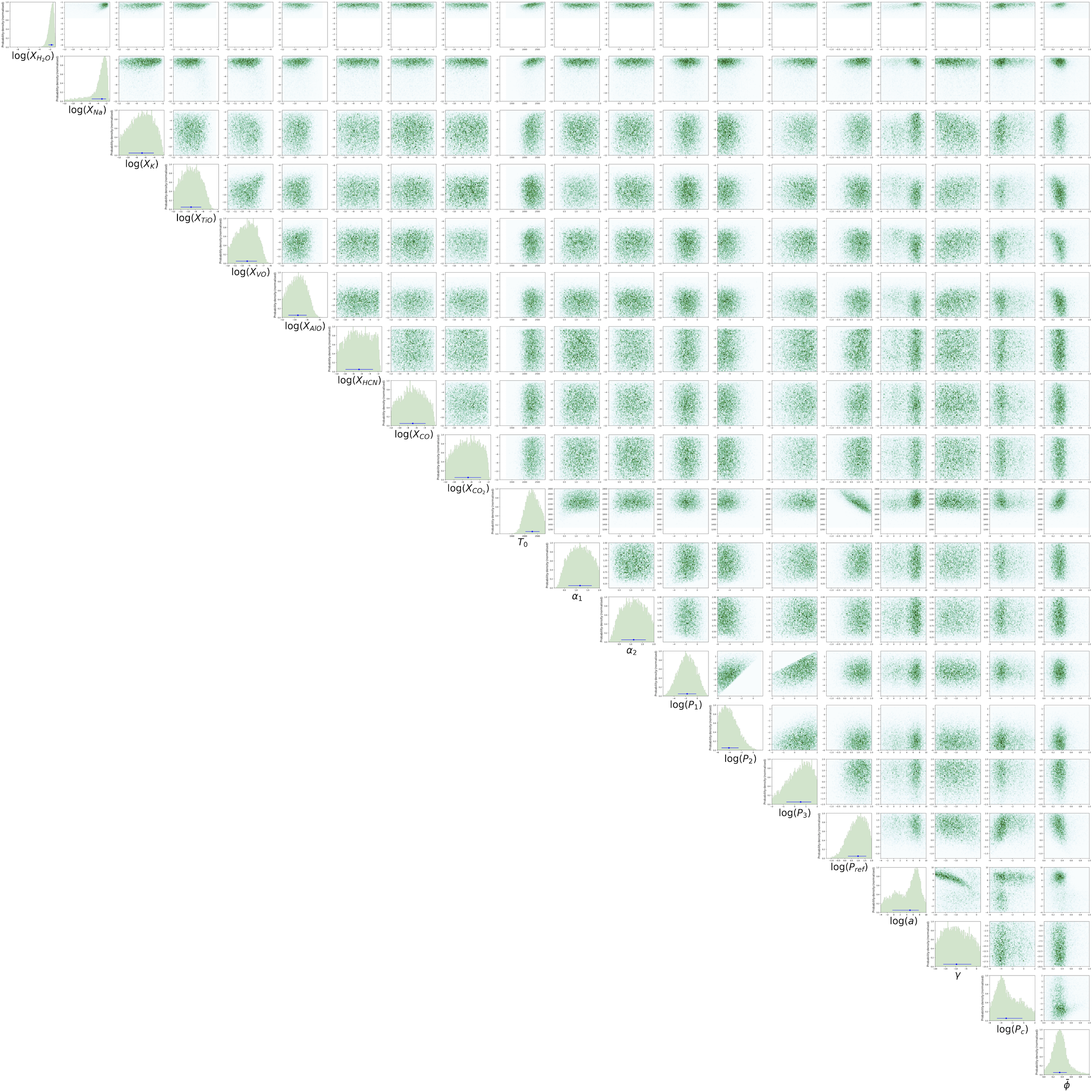}
 \caption{Marginalised posterior probability densities for the atmospheric retrieval performed using the AURA code on the full optical-infrared transmission spectrum of WASP-103b.}
\end{figure*}

\begin{figure*}
 \includegraphics[width=\textwidth]{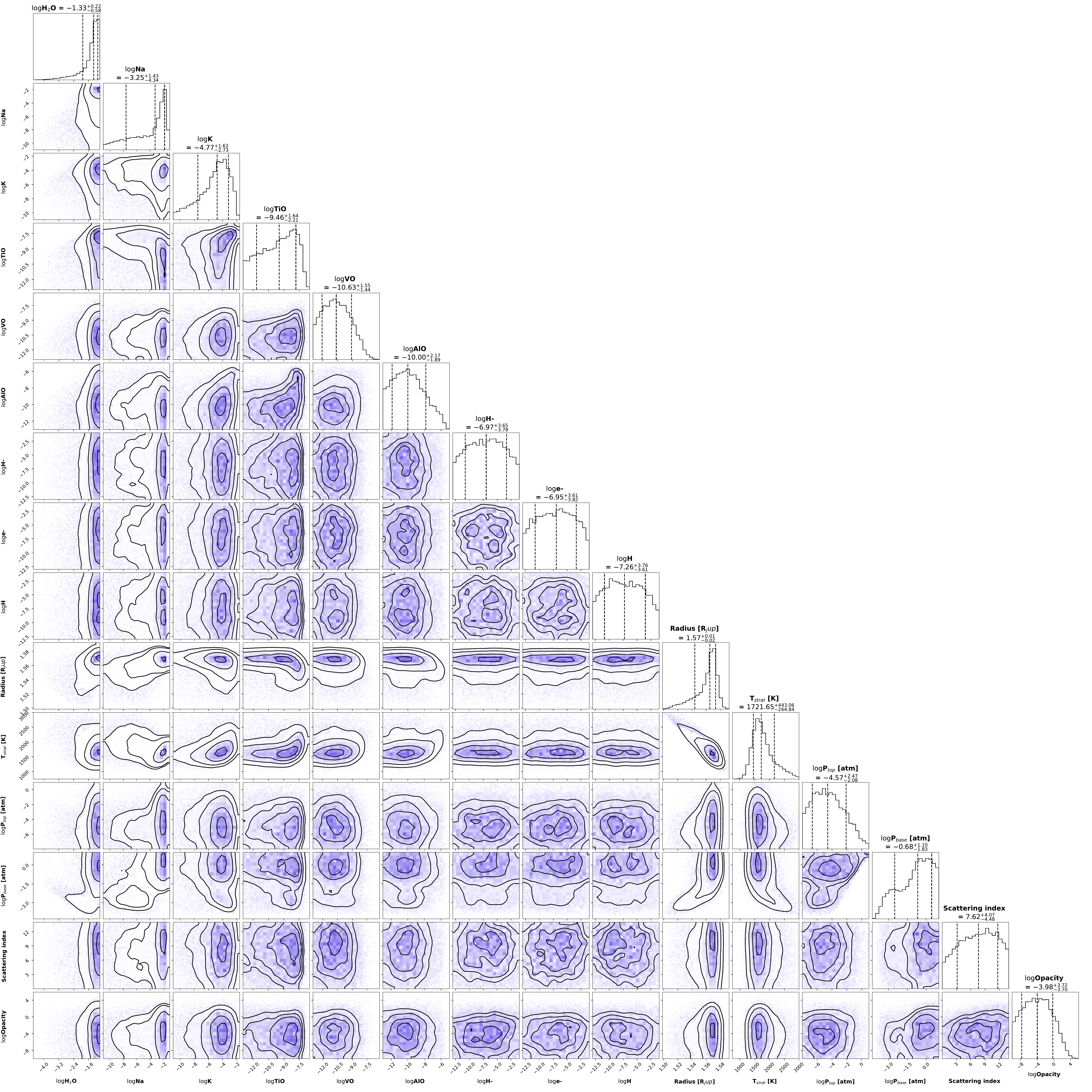}
 \caption{Marginalised posterior probability densities for the atmospheric retrieval performed using the NEMESIS code on the full optical-infrared transmission spectrum of WASP-103b.}
\end{figure*}

\bsp	
\label{lastpage}
\end{document}